\documentclass[sigconf]{acmart}
\usepackage{booktabs} 
\usepackage{hyperref}
\hypersetup{
    colorlinks=true,
    linkcolor=blue,
    filecolor=magenta,      
    urlcolor=cyan
    }
\usepackage{enumitem}
\usepackage{xcolor}
\setlist{leftmargin=5mm}

\AtBeginDocument{%
  \providecommand\BibTeX{{%
    \normalfont B\kern-0.5em{\scshape i\kern-0.25em b}\kern-0.8em\TeX}}}

\copyrightyear{2024}
\acmYear{2024}
\setcopyright{acmlicensed}
\acmConference[KDD '24] {Proceedings of the 30th ACM SIGKDD Conference on Knowledge Discovery and Data Mining }{August 25--29, 2024}{Barcelona, Spain.}
\acmBooktitle{Proceedings of the 30th ACM SIGKDD Conference on Knowledge Discovery and Data Mining (KDD '24), August 25--29, 2024, Barcelona, Spain}
\acmISBN{979-8-4007-0490-1/24/08}
\acmDOI{10.1145/3637528.3671462}
\begin{document}

\title{A Survey on Safe Multi-Modal Learning Systems}

\author{Tianyi Zhao}
\authornote{Both authors contributed equally to this research.}
\email{tzhao566@usc.edu}
\affiliation{%
  \institution{University of Southern California}
  \city{Los Angeles}
  \country{USA}}

\author{Liangliang Zhang}
\authornotemark[1]
\email{zhangl41@rpi.edu}
\affiliation{%
  \institution{Rensselaer Polytechnic Institute}
  \city{Troy}
  \country{USA}}

\author{Yao Ma}
\email{may13@rpi.edu}
\affiliation{%
  \institution{Rensselaer Polytechnic Institute}
  \city{Troy}
  \country{USA}}

\author{Lu Cheng}
\email{lucheng@uic.edu}
\affiliation{%
  \institution{University of Illinois Chicago}
  \city{Chicago}
  \country{USA}}

\begin{abstract}
    In the rapidly evolving landscape of artificial intelligence, multimodal learning systems (MMLS) have gained traction for their ability to process and integrate information from diverse modality inputs. Their expanding use in vital sectors such as healthcare has made safety assurance a critical concern.
    However, the absence of systematic research into their safety is a significant barrier to progress in this field. To bridge the gap, we present the first taxonomy that systematically categorizes and assesses MMLS safety. This taxonomy is structured around four fundamental pillars that are critical to ensuring the safety of MMLS: robustness, alignment, monitoring, and controllability. 
    Leveraging this taxonomy, we review existing methodologies, benchmarks, and the current state of research, while also pinpointing the principal limitations and gaps in knowledge. Finally, we discuss unique challenges in MMLS safety. In illuminating these challenges, we aim to pave the way for future research, proposing potential directions that could lead to significant advancements in the safety protocols of MMLS.
\end{abstract}






\maketitle

\section{Introduction}

Multimodal learning has rapidly evolved across various domains in the past few years. Compared to unimodal learning, it could more closely mimic human information processing, where multiple modalities, such as text, image, and audio, are incorporated to understand and interact with the world~\cite{baltruvsaitis2018multimodal,cheng2019xbully}.

Despite significant advancements in related research areas, deploying multimodal learning systems (MMLS) in real-world situations, especially in high-stakes environments like healthcare and autonomous driving, naturally brings safety challenges. These challenges are notably more intricate and acute than in unimodal environments \cite{han2022multimodal}.
Firstly, when implemented in real-world situations, MMLS are likely to face various types of uncommon events and risks. These include natural distribution shifts and deliberately crafted attacks intended to disrupt the systems. The complexity introduced by the multimodal setting may exacerbate these concerns, as it not only involves mismatches within each modality but also introduces domain shifts across modalities~\cite{mckinzie2023robustness}. Moreover, the rich information spanned by multiple modalities opens doors for cross-modal attacks by malicious entities, further increasing MMLS vulnerability~\cite{zhang2022towards}.
The growing risk of privacy is another significant concern.
The integration of multiple modalities enhances model efficacy but simultaneously provides attackers with additional information like modality alignment details~\cite{rao2023building},  exacerbating breaches of privacy. Furthermore, the increasing scales of MMLS further aggravate the challenge of pre-trained models inadvertently memorizing sensitive information from their training data~\cite{naseh2023understanding}.
MMLS are also at high risk of losing control and generating harmful responses such as discriminatory information or misinformation~\cite{schlarmann2023adversarial,li2024red}. Due to their increasing complexity, especially in large-scale MMLS, monitoring these systems becomes more challenging. This increased difficulty in oversight makes them more vulnerable to manipulation, resulting in a greater likelihood of performing harmful actions compared to their unimodal counterparts.

\begin{figure*}[]
  \centering
  \includegraphics[width=0.8\linewidth]{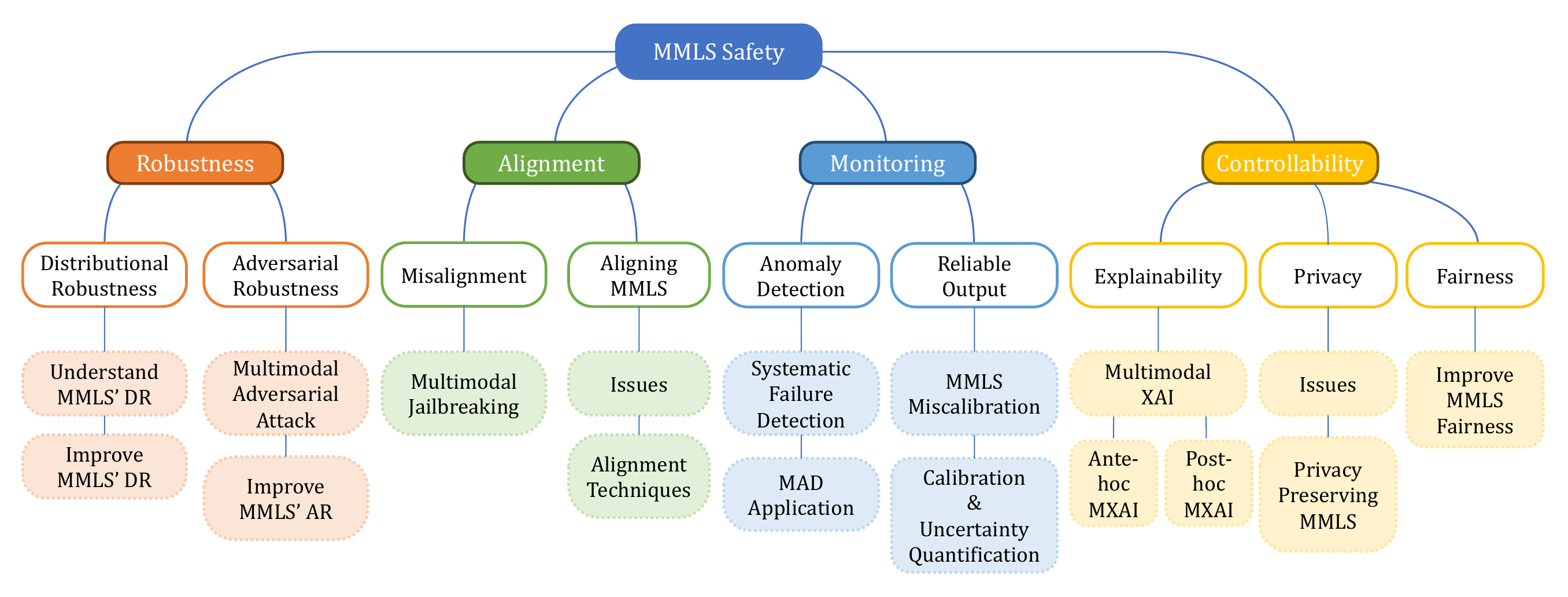}
  \caption{Taxonomy for Safety of Multimodal Learning Systems.}
  \label{fig:tax}
\end{figure*}

To advance the development of MMLS, it is essential to thoroughly consider these unique safety challenges and develop appropriate strategies. This survey serves to review recent advances in safe MMLS from four primary pillars: \textbf{robustness, alignment, monitoring,} and \textbf{controllability}.
For \textit{robustness}, we explore two main aspects: the resilience of MMLS against distribution shifts, and their robustness in the face of adversarial perturbations and attacks.
Regarding \textit{alignment}, we first examine the issue of the misalignment phenomenon in MMLS, noting their susceptibility to being jailbroken and producing harmful content. We then proceed to introduce the current techniques developed for aligning MMLS.
For \textit{monitoring}, our focus is twofold. First, we examine anomaly detection, which is crucial for promptly identifying potential hazards. Second, we investigate the reliability of model outputs to enhance the trustworthiness of MMLS.
\textit{Controllability} is divided into three subcategories. The first, interpretability, aims to study and increase the transparency and understandability of MMLS.
The second, privacy, discusses unique privacy challenges within MMLS, and various paradigms for preserving privacy in these contexts.
Lastly, fairness addresses inherent bias in these systems to ensure more equitable and fair responses.
We also discuss the limitations and summarize pertinent benchmarks related to these pillars and conclude the paper with open challenges.

In summary, our main contributions are \textbf{Taxonomy}: We present the first taxonomy (Figure \ref{fig:tax}) for MMLS safety; \textbf{Review}: Following this new taxonomy, we conduct a comprehensive review of existing works, highlighting their limitations; and \textbf{Open Challenges}: We identify some unique challenges related to the safety of MMLS and suggest potential prospects.

\section{Robustness}

Robustness emerges as a critical research problem when addressing safety concerns in MMLS.
In this section, we delineate the problem along two dimensions: \textit{robustness against distributional shift}, referring to the ability of MMLS to adapt to unexpected natural variations;
and \textit{adversarial robustness}, denoting their resilience to deliberately crafted threats.
\subsection{Robustness against Distributional Shift}
A standard assumption in ML is consistent distribution, but real-world scenarios often show mismatches between training and testing data. In MMLS settings, these mismatches are more complex, involving shifts within and across modalities.
Understanding and enhancing MMLS robustness against distribution shifts is crucial for their safe deployment.

\subsubsection{Understanding distributional robustness in MMLS}
Most prior research on robust MMLS focuses on vision-language models (VLMs), which are susceptible to distribution shifts in both image and text modalities.
Yet they generally show greater robustness than unimodal models~\cite{fang2022data}. Understanding the reasons behind this robustness gain is vital for advancing MMLS safety.
For instance, \cite{fang2022data} identifies diverse image training data as the key to CLIP's robustness, excluding language supervision and contrastive loss function. However, contrasting views exist~\cite{xue2023understanding}, suggesting these latter factors also make contributions. Consensus on these findings is yet to be established.
Modality mismatch represents a unique distribution shift in MMLS, which occurs when the data modalities during testing differ from those in training. 
Earlier studies primarily explored the impact of missing modalities on MMLS's robustness~\cite{ma2022multimodal}. 
More recent research starts considering more generalizable settings~\cite{mckinzie2023robustness}, investigating the effects of both missing and newly added modalities during test time on MMLS robustness.
\subsubsection{Improving MMLS distributional robustness}

Current efforts to improve the distributional robustness of MMLS often involve \textit{data augmentation} techniques and \textit{robust training} strategies.

\noindent\underline{\textsl{Data Augmentation.}}
Data augmentation techniques, proven effective in unimodal systems for enhancing model robustness, are naturally extended to MMLS.
However, applying unimodal strategies independently to each modality risks losing semantic coherence~\cite{liu2022learning}.
Consequently, multimodal techniques preserving modality consistency while improving robustness have emerged. For example, RobustMixGen for VLMs uses a pre-synthesis data separation module to separate data into objects and backgrounds. It maintains cross-modal semantics and boosts model robustness by reducing spurious correlations~\cite{kim2023robustmixgen}.

\noindent\underline{\textsl{Robust Training.}}
Research in robust training strategies is diverse and spans from improving pre-training strategies to adaptation and fine-tuning methods. 
For instance, a multi-task optimization algorithm has been developed to automatically identify the optimal data fusion strategies, enhancing models' robustness against cross-modal distribution shifts~\cite{ma2022multimodal}.
Andonian et al.\cite{andonian2022robust} presents a novel contrastive learning method for pretraining, combining knowledge distillation and soft alignments to improve robustness. Other studies include devising robust adaptation methods. For instance, \cite{wortsman2022model} shows that fusing weights from zero-shot and fine-tuned CLIPs enhances distributional robustness, with further improvements observed through weight averaging from CLIPs fine-tuned under different hyperparameters. 

\subsection{Adversarial Robustness}
MMLS robustness to adversarial attacks faces unique challenges. Malicious users might use the information from multiple modalities and cross-modal adversarial attacks, such as subtle alterations in an image affecting text predictions, further complicate the issue~\cite{zhang2022towards}.

\subsubsection{Multimodal adversarial attack.}
Multimodal adversarial attack aims to trick a target MMLS by subtly altering its multimodal inputs.
The objective function for multimodal adversarial attacks can generally be formulated as~\cite{tian2021can}:
\begin{equation}
\mathop{\arg\max}_{\{x_i^{adv}\}_{i=1}^n} \mathcal{L}(\{x_i^{adv}\}_{i=1}^n,y ; \theta)\quad s.t.\,||x_i^{adv}-x_i||_p\le\epsilon_i,
\end{equation}
where $\{x_i\}_{i=1}^n$ denotes the input composed of $n$ modalities, $\{x_i^{adv}\}_{i=1}^n$ are the generated adversarial samples, and $\epsilon_i$ is the perturbation budget for the $i$-th modality. 
The attacker optimizes $\mathcal{L}$ parameterized by $\theta$ to induce model mispredictions within permissible budgets.
An effective multimodal adversarial attack needs to properly address the modality gap between multiple input data modalities.
\cite{zhang2022towards} studies multimodal attacks on VLMs, discovering that launching unimodal attacks independently often fails. They then propose Co-Attack, which coordinates image and text attacks through a sequential strategy 
Another challenge lies in executing effective, downstream-agnostic attacks, especially within the prevalent `pretrain-then-finetune' framework for MMLS, as designing specific attacks for a pre-trained encoder targeting each task is impractical.
Pioneering work by \cite{zhou2023advclip} proposes a method for generating broadly applicable adversarial noise, ensuring adversarial samples' significant deviation from decision boundary in the feature space via a topology-deviation-based generative adversarial network.

\subsubsection{Enhance adversarial robustness of multimodal learning models.}
Current works on improving the adversarial robustness of MMLS center around \textit{robust fusion strategy} and \textit{adversarial training}. 

\noindent\underline{\textsl{Robust Fusion and Alignment.}}
The unique modality alignment process of MMLS, compared to unimodal models, has catalyzed extensive research into enhancing their adversarial robustness.
For instance, \cite{yang2022defending} 
improves MMLS single-source adversarial robustness by designing a fusion strategy that detects inconsistencies across modalities and executes feature fusion to counteract the perturbed modality.
A constrastive learning framework has been proposed that captures both intra- and inter-modality dynamics, filtering out noise present in various modalities and thereby yielding more robust representations~\cite{lin2022multimodal}.

\noindent\underline{\textsl{Adversarial Training.}}
Adversarial training enhances robustness effectively, yet its application in MMLS presents challenges, including higher computational demands and scalability issues.
To approach these, \cite{gan2020large} adopts the free adversarial training strategy that obtains parameters' gradients with almost no extra computational expense.
Another challenge lies in improving the generalization of adversarial training procedures within the emerging pretrain-then-finetune paradigm for MMLS.
An adversarial pre-training strategy for VLMs is introduced in \cite{gan2020large}, focusing on adversarial training in the embedding space of each modality rather than directly perturbing image pixels or text tokens, effectively enhancing model performance in a task-agnostic manner. 

\subsection{Benchmark} 
The assessment of model distributional robustness commonly employs metrics such as \textit{Effective Robustness}, along with \textit{Relative Robustness} and the \textit{MultiModal Impact Score} to measure average performance drops during distributional shifts~\cite{fang2022data,qiu2022multimodal}.
Systematic benchmarks are constantly being proposed as well. 
For instance, \cite{li2020closer} integrates nine VQA datasets to study four generic robustness in VLMs.
Recently, benchmarks for large MMLS have emerged, such as the one introduced in \cite{han2023well}. This benchmark assesses GPT-4V's~\cite{achiam2023gpt} zero-shot generalization on 13 varied datasets, offering insights into its adaptability to distribution shifts.
\subsection{Discussion}
Despite the ongoing research, a clear understanding of what underpins the robustness of MMLS remains under-explored, highlighting the need for further studies. Current studies tend to concentrate on VLMs, suggesting the necessity of expanding research focus on a broader spectrum of MMLS.
Besides, MMLS vulnerability to sophisticated real-world attacks underscores the importance of enhancing their adversarial robustness.
This generally involves devising potent attacks and corresponding defenses, both of which must navigate the complexities of cross-modal information interplay in MMLS.
Another challenge is ensuring these defenses can be adapted to the unpredictable nature of real-world attacks.

\section{Alignment}
Aligning MMLS with human values is a crucial yet challenging aspect of safe deployment.
Recent studies have highlighted vulnerabilities in current foundation models, revealing that they can be manipulated to yield harmful and misaligned outputs. This issue could be further exacerbated in MMLS than unimodal foundation models such as LLMs.
Therefore, delving into this phenomenon and developing more safely-aligned MMLS is a critical area of interest.
We discuss this issue from two perspectives: the misalignment in MMLS and strategies for their alignment.

\subsection{Misalignment: Jailbreaking MMLS}
The development of foundation models introduces a safety challenge termed `jailbreak', where malicious users attempt to circumvent safeguards and instruct the model to produce harmful responses that are misaligned with ethical standards.
Previous research has mainly focused on unimodal foundation models, particularly LLMs, with studies on MMLS being relatively new and concentrating largely on VLMs.
One of the most prevalent jailbreak approaches is \textit{adversarial prompting}.
Compared to LLMs, MMLS can be more vulnerable to jailbreak attacks owing to the integration of additional modalities. 
For instance, \cite{carlini2024aligned} successfully jailbreaks VLMs via constructing adversarial image input.
Besides, \cite{qi2023visual} designs a simple attack that maximizes the generation probability of harmful responses, effectively jailbreaking VLMs with a single visual adversarial example. 
More recent work~\cite{niu2024jailbreaking} investigates the potential of constructing adversarial textual input based on image jailbreak prompt, introducing new approaches to jailbreak MMLS.

\subsection{Aligning MMLS}
At present, the two most common techniques for achieving model alignment are Instruction Tuning~\cite{liu2024visual}, which finetunes models using collections of pairs of instructions and faithful outputs, and Reinforcement Learning from Human Feedback (RLHF), which employs reinforcement learning techniques to optimize model outputs~\cite{stiennon2020learning,ouyang2022training}.
When aligning MMLS, a significant hurdle is the scarcity of available data. 
Multimodal data pairs are more challenging to gather from the internet compared to textual data alone. Another issue is the quality of the data collected. For example, an image paired with a brief description is not as informative as one accompanied by a comprehensive narrative.
LLaVA-RLHF~\cite{sun2023aligning} marks a pioneering effort in adapting RLHF from purely textual applications to vision-language alignment tasks, using human annotated preferences to fine-tune multimodal models via reinforcement learning. It effectively addresses multimodal misalignment at a relatively low cost for annotations.
\cite{li2023silkie} first constructs a high-quality instruction set based on various multimodal instruction tuning sources and AI annotations, and then employs direct preference optimization~\cite{rafailov2024direct} to enhance VLMs' alignment, resulting in more faithful systems and mitigating misalignment behaviors.

\subsection{Benchmark}
Common evaluation metrics include \textit{Attack Success Rate, Recognition Success Rate, }and \textit{Defense Success Rate}~\cite{wu2023jailbreaking}.
Recently, researchers have begun to establish systematic benchmarks for measuring misalignment in MMLS.
For instance, \cite{li2024red} proposes a multimodal dataset RTVLM for a comprehensive assessment of VLMs across various red-teaming cases.
\cite{niu2024jailbreaking} expands upon AdvBench to construct a multimodal benchmark, AdvBench-M, which consists of eight distinct categories of harmful behaviors for evaluation.

\subsection{Discussion}
Aligning multimodal foundation models is notably more complex because of the involvement of multiple modalities~\cite{qi2023visual}. This complexity increases the potential for vulnerabilities and intensifies the challenge of establishing robust defenses. 
For instance, incorporating visual data into LLMs creates opportunities for adversarial attacks, which can be executed more readily than with purely discrete textual inputs.  
Consequently, this facilitates the jailbreaking of MMLS, often leading to the generation of toxic content, misinformation, and hallucination.
Although researchers have devised various strategies for aligning MMLS, the effectiveness of these methods is still under scrutiny. Even with attempts at safety alignment, MMLS can be compromised with relative ease. Thus, the pursuit of alignment within MMLS is fraught with difficulties and remains an area in need of more thorough exploration and understanding.

\section{Monitoring-Anomaly Detection}
Monitoring is crucial for AI safety. Here, we discuss an important step for monitoring system behavior: detecting anomalies. Anomaly detection (AD) refers to the problem of finding patterns in data that do not conform to expected behavior. While AD has been extensively discussed in various fields, MMLS introduce extra complexity owing to heterogeneous modality information, making Multimodal Anomaly Detection (MAD) more challenging. Our review of MAD includes MAD assumptions, research techniques, and benchmarks.

\subsection{Failure Detection in MAD}
A common practice of MAD in MMLS is to detect failure, which can serve as an early warning system for potential failures by identifying irregularities. Deployment failures of MMLS are prevalent despite rigorous testing before development. One of the challenges lies in anticipating and testing all potential failures in advance. 
To bridge this gap, MULTIMON \cite{tong2024mass} is introduced to automatically detect systematic failures in deployed MMLS. It first identifies individual failures based on CLIP embeddings, which encode multimodal information, and erroneous agreement, and then leverages language models to describe patterns of failures.
The framework proposed by~\cite{jain2022distilling} aims to detect failure modes within the latent space. It seeks to discover a hyperplane that optimally distinguishes between correct and incorrect examples, interprets the patterns of errors, and then enhances model reliability during training along the axes of failure.
Similarly, DOMINO \cite{eyuboglu2022domino} tackles the challenge of identifying underperforming subsets within MMLS, particularly dealing with audiovisual inputs. Recent studies \cite{d2022spotlight,yeh2020completeness} have presented Slice Discovery Methods (SDMs) that use model representations to detect underperforming segments, however, these methods often lack a robust framework for quantitative analysis. Thus DOMINO has been developed as an extensive evaluation framework for SDMs in various contexts and introduced as an innovative SDM that employs cross-modal embeddings and an error-aware mixture model for improved performance. It outperforms prior methods, demonstrating improved coherence and performance in slice discovery.


\subsection{Other Applications of MAD}
\subsubsection{Vision-Language MAD}
Vision modality faces sample variations due to factors like color and light, causing latent space sparsity and disrupted correlations, which impact task performance. Conversely, the more uniform language modality often performs better. Existing research \cite{pang2021deep} mostly focuses on unimodality, particularly vision, overlooking multimodal benefits. Meanwhile integrating multimodalities also introuces challenges to be addressed.

\noindent\underline{\textsl{Zero-shot VLMs AD.}}
Motivated by the challenges posed by the long-tailed distribution and privacy concerns in AD data, zero-shot AD emerges, offering versatility across various AD tasks. The same has been observed in MAD. VLMs like CLIP acquire expressive representations through extensive training in vision-language annotated pairs. Based on this, text prompts can be used to extract knowledge from the models without additional fine-tuning, enabling zero-/few-shot transfer to downstream tasks.
Given image-text pairs $\{(x_t,s_t)\}_{t=1}^T$, CLIP trains a text encoder $g$ and a visual encoder $f$ 
to maximize the correlation between them in terms of cosine similarity $<f(x),g(s)>$. Given an input image $x$ and a set of text $S$, CLIP can predict the probability of $x$ belonging to class $c$ as follows:
\begin{equation}
\begin{split}
P(f(x),g(s_i)) =\frac{\exp(<f(x),g(s_i)>/ \tau)}{\sum_{s\in S}\exp(<f(x),g(s_i)>/ \tau)},
\end{split}
\label{eq-CLIP}
\end{equation}
where $\tau>0$ is a hyperparameter. In CLIP, the text prompt template commonly looks like ``\texttt{a photo of a [c]}'', where \texttt{[c]} represents the target class name. It differs from vision tasks that involve many objects and use objects' names as \texttt{[c]}. 
WinCLIP \cite{jeong2023winclip} employs anomaly text prompts and calculates the corresponding scores using Eq.\ref{eq-CLIP}. For prompt optimization, a linear combination of the focal loss \cite{lin2017focal} and dice loss \cite{milletari2016v} is usually used. Subsequent research mainly follows this approach to compute anomaly scores \cite{zhou2023anomalyclip,deng2023anovl,chen2023clip}.

\noindent\underline{\textsl{Adding high-quality anomaly description.}}
Most existing MAD methods only provide anomaly scores and require manual thresholds to differentiate between normal and abnormal samples, limiting wider use. Incorporating high-quality anomaly descriptions can enhance understanding and facilitate practical applications. AnomalyGPT \cite{gu2023anomalygpt} designs an MMLS-based image decoder for anomaly mapping and employs prompt embedding for fine-tuning. It supports multi-turn dialogues and exhibits impressive few-shot in-context learning capabilities. Yet, it underutilizes MMLS's vision comprehension potential.
Myriad~\cite{li2023myriad} addresses this by skillfully incorporating domain knowledge into embeddings, aligning vision-language features, and introducing an Expert Perception module. This module embeds prior knowledge from vision experts as tokens intelligible to LLMs, ensuring precise AD with detailed descriptions, including the anomaly location, type, and explanation of the judgment.

\subsubsection{Video Surveillance MAD}
Another common multimodal data is video. Video Surveillance \cite{pang2021deep} plays a crucial role in monitoring areas for security. It is often treated as an AD problem due to the substantial amount of unlabeled data. Meanwhile, multimodal information in video frames is often underutilized, indicating the need for more effective use.
Cutting-edge video MAD methods involve incorporating multimodal signals. For example, MSAF \cite{wei2022msaf} integrates RGB, optical-flow, and audio modalities, transforming video-level ground truth into pseudo clip-level labels for subsequent training, thereby aligning diverse information to enhance learning. 
Unlike current works that directly use extracted features for frame-level classification, VadCLIP~\cite{wu2023vadclip} devises an approach that leverages the frozen CLIP and multimodal associations between video features and fine-grained language-image alignment information.
Video-AnomalyCLIP~\cite{zanella2023delving} further explore the potential of CLIP, effectively learning text-driven directions for abnormal events by manipulating the latent feature space and projecting anomalous frames onto these directions to exhibit large feature magnitudes for a particular class.

 

\subsection{Benchmark}
In failure detection, datasets typically consist of normal data without explicitly labeled anomaly subsets. For failure detection in MAD, datasets vary widely, covering areas from medical imaging to image-text corpora. Conversely, in AD scenarios, datasets are specifically annotated to highlight anomalies. When evaluating Vision-Language MAD, commonly used benchmark datasets include MVTec-AD \cite{bergmann2019mvtec}, VisA \cite{zou2022spot}, KSDD2 \cite{bovzivc2021mixed}, MPDD \cite{jezek2021deep}, MTD \cite{huang2020surface}, 
etc. These industrial inspection datasets are annotated at the pixel level and cover a wide range of object subsets. Datasets for Video Surveillance MAD include ShanghaiTech \cite{liu2018future}, UCF-Crime \cite{sultani2018real} and  XD-Violence \cite{wu2020not}, an audiovisual abnormal dataset. Performance evaluation is primarily based on metrics such as the area under the ROC curve, area under the precision-recall curve, and F1 score. Other metrics include Per-Region Overlap scores, which assess anomaly segmentation performance, and Average Precision.

\subsection{Discussion}
Existing MAD works primarily center around CLIP, resulting in limited exploration of broader multimodal applications \cite{yoo2021multimodal,wang2023multimodal,ji2020multi}. 
Moreover, current efforts predominantly focus on addressing unsupervised or weakly-supervised MAD challenges, while overlooking the issue of handling incomplete and inaccurate anomaly labels. 
Notably, limited efforts have been put into studying supervised MAD, where labeling costs might be high. 
Additionally, while AD has been applied to diverse scenarios like financial fraud and medical systems, the applications of MAD remain limited.
\section{Monitoring-Reliable Model Outputs}

Ensuring MMLS's safe deployment requires establishing monitoring mechanisms to assess model reliability. Existing ML methods face challenges in accurately conveying their understanding and providing reliable responses.
In MMLS, these issues may become more pronounced given their increased complexity.
We discuss this problem from two perspectives: firstly, the phenomenon of miscalibration in MMLS, and secondly, the strategies for achieving calibration and quantifying uncertainty within these systems.


\subsection{Miscalibration in MMLS}
Miscalibrations in AI systems can manifest as overconfidence, where models inaccurately assign high probabilities to incorrect predictions, or underconfidence, where models fail to recognize the accuracy of their correct predictions with sufficient certainty.
Similar to uni-modal learning, miscalibration is a prevalent issue in MMLS yet remains underexplored.
Despite the involvement of various modalities, MMLS may exhibit a bias towards relying on a single modality for decision-making, resulting in the issue of over-confidence \cite{ma2023calibrating}.
Recently, few pilot studies have emerged focusing on the foundation MMLS. For example, concerns about miscalibration have been raised regarding zero-shot VLMs such as CLIP. \cite{oh2023towards} conducts further studies on fine-tuned VLMs, uncovering that traditional fine-tuning strategies can significantly affect model calibration.
Ensuring MMLS are well-calibrated is vital for their safe use, with research concentrating on model calibration and reliable uncertainty estimation (specifically when models are not well-calibrated).

\subsection{Calibrating and Estimating Uncertainty}
A well-calibrated model should make predictions with high uncertainty when its accuracy is in doubt, and conversely, it should predict with confidence when it is likely to be accurate.
The diversity of modalities in MMLS presents both challenges and potential breakthroughs for the development of new approaches for calibrating MMLS. 
For instance, a regularization-based technique CML is proposed for better calibration of MMLS, which calibrates the relationship between confidence and the number of modalities involved in multimodal learning~\cite{ma2023calibrating}.
Recent uncertainty estimation research primarily focuses on Bayesian and Conformal Prediction (CP) methods, with extensive work in unimodal settings (e.g., \cite{zhao2024link,su2024api}) but limited studies in multimodal contexts.
Bayesian-based approaches place a distribution over
model parameters and employ marginalization for predictive
distribution. There are generally two types of predictive uncertainty: aleatoric uncertainty and epistemic uncertainty. 
\cite{zhang2023provable} pioneers the application of Dempster-Shafer theory~\cite{shafer1976mathematical} to model epistemic uncertainty for each modality and introduces an uncertainty-aware weighting method for dynamic modality fusion. Meanwhile, Subedar et al.~\cite{subedar2019uncertainty} introduce a Bayesian framework for audiovisual applications, quantifying modality-wise epistemic uncertainty and deepening Bayesian DNNs by combining deterministic and variational layers.
While Bayesian-based approaches are widely used, their computational intensity renders them resource-heavy. 
Conversely, CP-based methods~\cite{shafer2008tutorial}
have emerged as scalable, efficient, and model-agnostic alternatives for uncertainty estimation. CP yields statistically valid prediction sets or intervals from any black box model, with the sole assumption of data exchangeability.
Currently, the application of CP in MMLS is limited. \cite{dutta2023estimating} proposes a CP procedure for foundation models using multimodal web data. It generates plausibility scores based on modality alignment, and then uses them for Monte Carlo-based CP procedure.
\vspace{-3mm}
\subsection{Benchmark}
To measure the reliability of model outputs, a commonly used metric is accuracy vs uncertainty (AvU)~\cite{krishnan2020improving}, which considers four types of predictions: accurate and certain, inaccurate
and uncertain, accurate and uncertain, inaccurate and certain.
Other measurements include Expected Calibration Error (ECE)~\cite{guo2017calibration}, Uncertainty Calibration Error (UCE)~\cite{laves2019well}, reliability diagrams~\cite{hartmann2002confidence}, and Bayesian active learning by disagreement (BALD)~\cite{houlsby2011bayesian}. Datasets used for benchmarking can differ widely depending on the task and method.
\vspace{-6mm}
\subsection{Discussion}
Advancing MMLS's reliable output stands as a critical yet presently inadequately investigated area of research.
To improve reliability, it is necessary to address the prevalent issues of miscalibration that these systems often exhibit. 
One of the major challenges in this endeavor involves tackling the intricate interactions between different modalities.
Although this complexity and variance contribute to miscalibration, it also presents opportunities for effective calibration and precise uncertainty estimation.
Current multimodal calibration and uncertainty quantification primarily center around Bayesian-based approaches, which tend to be computationally intensive. The emerging CP-based method offers a scalable and model-agnostic alternative. However, its application in MMLS is still in its infancy, highlighting a ripe area for further investigation.

\section{Controllability - {Explainability}}
Being able to comprehend and explain the decisions made by the system is critical for the controllability of MMLS. The complexity of MMLS makes it difficult to interpret, opaque, and black-box models with little or no understanding of their internal states and decision-making process \cite{joshi2021review}. Gaining meaningful knowledge and understanding of how and why the model arrived at a particular decision or outcome is crucial in model explainability, making it one of the important evaluation metrics
In this section, we offer a concise overview of Multimodal Explainable Artificial Intelligence, categorizing MXAI into ante-hoc and post-hoc approaches. A comprehensive review of MXAI can be found in \cite{rodis2023multimodal}
\subsection{Multimodal XAI}

\subsubsection{Ante-hoc MXAI} 
Ante-hoc methods integrate interpretability modules directly into the primary model for simultaneous learning.
In multimodal contexts, emphasizing the modeling of cross-modal interactions and assessing modalities' relevance is essential.
For example, \cite{li2018vqa} introduces a multitask learning approach for VQA that concurrently trains an answer predictor and a textual explanation generator, leveraging an attention mechanism for cross-modal dynamics. 
Recently, research on multimodal foundation models has emerged.
\cite{wang2023towards} devises an interactive multi-agent VQA framework based on VLMs. 
This framework consists of a seeker and responder that actively generates and exchanges information and candidate responses related to the query, and an integrator that synthesizes candidates and generated hypotheses from the seeker and responder to formulate the final answer.

\subsubsection{Post-hoc MXAI} 
Post-hoc methods generally refer to training an interpretability module based on the output of the primary model.
To address the fundamental challenges of capturing interactions across modalities, DIME \cite{lyu2022dime} proposes to disentangle the black-box model into unimodal contributions and multimodal interactions. This allows for fine-grained analysis of multimodal models while maintaining generality across arbitrary modalities, model architectures, and tasks. 
Another post-hoc method \cite{panousis2024discover} focuses on understanding individual neuron functionality in tasks, utilizing VLMs and unique network layers to achieve high activation sparsity. It enables the generation of simple textual descriptions, facilitating a direct investigation of the network’s decision process.

\subsection{Benchmark}
MXAI datasets are typically organized based on tasks, I/O types, dates, and brief descriptions.
In evaluating MXAI, the focus is on assessing the effectiveness of explanations in elucidating model decisions. Evaluation methods can be categorized into three types: application-grounded, involving domain experts for real-world utility; human-grounded, testing understandability with individuals regardless of expertise; and functionally-grounded, using formal criteria to assess explanation quality. Detailed descriptions of MXAI benchmarks can be found in \cite{rodis2023multimodal}. 

\subsection{Discussion}
Formalizing definitions and terminology is crucial for the MXAI field's development, as current informal practices and inconsistent definitions hinder progress.
Specifically, it is crucial to define explanations precisely and establish standardization for measuring MXAI's efficiency to enable comparative evaluation of different methods. 
Another significant challenge arises from the absence of ground-truth modality explanations. For instance, in contrast to textual explanations, the lack of corresponding ground-truth visual explanation data hampers the development of related explanation methods, especially in supervised learning scenarios.

\section{Controllability - Fairness}
MMLS incorporates data from various sources, presenting unique fairness challenges. Prior work indicates that biases in MMLS stem from individual modalities, which could then be amplified during later modality fusion process~\cite{yan2020mitigating}. It is crucial to identify and address the specific fairness challenges and underlying causes in MMLS to effectively mitigate discrimination.

\subsection{Improving Fairness in MMLS}
Addressing fairness in multimodal scenarios is more challenging compared to unimodal due to cross-modality correlation. 
Following categorization in uni-modal fair machine learning \cite{cheng2021socially}, we organize fairness algorithms in MMLS into three groups: pre-processing, in-processing, and post-processing.
\subsubsection{Pre-processing}
Pre-processing techniques operate on original input data to remove its inherent bias.
For instance, In the context of Automated Video Interviews, Booth et al. \cite{booth2021bias} investigate and tackle gender bias by analyzing verbal, paraverbal, and visual features.
They present two strategies to mitigate gender-related bias within the feature set: first, gender-norming, which involves separately normalizing the features of men and women before training, and second, iterative gendered predictor removal, a process that systematically eliminates the predictors most indicative of gender information.
This study also indicates that combining multiple modalities without addressing biases hardly boosts accuracy and tends to amplify unfairness compared to unimodal models. Nonetheless, bias mitigation techniques like feature reduction \cite{schmitz2022bias} can achieve an admissible fairness-accuracy trade-off.
\subsubsection{In-processing}
In-processing techniques involve devising training algorithms to mitigate discrimination. Compared to the pre-processing techniques, in-processing methods tend to achieve superior trade-off between fairness and task performance.
For instance, FairCVtest \cite{pena2020bias} offers a fictional recruitment testbed demonstrating how biases can be affected by unimodal factors. It further ensures fair multimodal outcomes by generating agnostic representations based on \textit{SensitiveNets} learning strategy. 
However, addressing unfairness in MMLS involves more than just tackling unimodal biases. FMMRec \cite{chen2023fmmrec} finds that multimodal recommend systems' bias stems from repetitively identifying sensitive information across modalities.
Adversarial learning method is then employed to disentangle and filter out sensitive information from modality representations.
Other approaches involve integrating the aforementioned techniques. 
For instance, \cite{yan2020mitigating} mitigates gender and race biases through a combination of adversarial learning and data balancing, employing both in-processing and post-processing strategies.
\subsubsection{Post-processing}
Post-processing techniques are performed after training by accessing a holdout set not involved during training.
For instance, \cite{chen2020exploring} determines group-specific random thresholds based on the intersection of group-specific ROC curves. By adding this to the clinical MMLS outputs, the prediction bias can be removed.
Notably, post-processing methods are exclusively usable in black-box settings where the model and training data are inaccessible.
Although pre-processing and post-processing are more generalizable, easier to implement, and less time-consuming, they may lead to uncertain outcomes if not customized for specific algorithms\cite{mehrabi2021survey}.

\subsection{Benchmark}
Datasets crucial for ensuring fairness in MMLS are limited due to privacy concerns and the intricate nature of multimodal data. Given the scarcity of research in the field, we highlight several representative datasets for reference.
MovieKens \cite{ni2023content} is a widely used multimodal benchmark dataset for movie recommendation. IEMOCAP \cite{busso2008iemocap} is a emotion recognition dataset covering audio, text, and video. Additionally, the clinical database MIMIC-III \cite{johnson2016mimic} and the image-text corpus Multi30K \cite{elliott2016multi30k} also contribute to fairness research in clinical data and multilingual contexts, respectively. 
Notions of fairness vary in different systems. Various fairness notions (e.g., \textit{Equalized Odds} and \textit{Statistical Parity} \cite{mehrabi2021survey}) in MMLS draw inspiration from concepts established in unimodal settings. Key fairness categories encompass individual fairness, focusing on individual samples, and group fairness, targeting protected groups such as Black or Female \cite{chen2020exploring}. In the context of multilingual MMLS, additional fairness metrics like \textit{Multilingual Individual} and \textit{Multilingual Group Fairness} have been introduced \cite{wang2021assessing}. The former necessitates similar semantics across languages, linking text to images, while the latter measures equalized predictive performance across languages.

\subsection{Discussions}
In contrast to the abundance of datasets available for fairness studies in unimodal scenarios, there is a notable absence of reliable datasets for fairness studies in MMLS. This gap is attributed to the intricate nature of multi-modal information, privacy concerns, and the diverse system considerations related to fairness. Existing work, such as \cite{schmitz2022bias}, compares biases introduced by different modalities and their combinations, yet falls short of delving into the fundamental causes of multimodal unfairness and proposing effective solutions. One may explore the causes from an uncertainty perspective \cite{tahir2023fairness}. Furthermore, there is a crucial need for the introduction of new definitions of fairness. While some notable work has suggested some \cite{wang2021assessing}, it lacks a thorough explanation of why this new definition is necessary or when it should be applied.

\section{Controllability - Privacy}

\subsection{Privacy Issues in MMLS}


Significant privacy concerns regarding MMLS have emerged due to their high risk of memorizing and leaking private information \cite{chen2023can,naseh2023understanding}.
Membership Inference Attacks~\cite{ko2023practical}, for example, could reveal an individual's health status by checking if their record is in a multimodal disease database.
Besides, reconstruction attacks may exploit MMLS's memorized information to extract private data, such as identifying residents in a geospatial system using modality alignment information between street images and resident details~\cite{rao2023building}.
The recent emerging large MMLS following the pretrain-then-finetune paradigm further intensifies privacy concerns. Each stage of these models, pre-training, fine-tuning, and prompt interaction, carries a high risk of privacy leakage~\cite{rao2023building}.

\subsection{Preserving Privacy in MMLS}

\subsubsection{Differentially private training.}
The differential privacy (DP) technique, recognized as a gold standard in privacy preservation, ensures data protection by introducing randomness into its computational processes.
DP in MMLS remains understudied.
Multimodal DP was first introduced to address the heightened risk of privacy leakage arising from MMLS memorizing sensitive training data information~\cite{huang2023safeguarding}.
In this study, DP-CLIP is proposed to safeguard vision-language tasks by learning differentially private image and text representations.
It employs per-batch, instead of per-sample, noisy stochastic gradient descent to train the encoder, considering that the CLIP loss requires contrasting data from different pairs.

\subsubsection{Machine unlearning.}
Machine unlearning aims at selectively erasing information (e.g., sensitive information) from a trained model without the need for complete retraining.
Progress in machine unlearning has advanced in unimodal learning, but its application in MMLS is still in its infancy.
Compared to unimodal scenarios, the key challenge lies in effectively addressing the inherent interdependencies across modalities.
One work~\cite{shaik2023framu} presents a modality-specific attention mechanism that generates individual attention scores per modality, averages them across modalities for each data point to determine its final score, and then decides on data removal and model parameter updates based on this score. 
Recently a more comprehensive framework for multimodal unlearning has been proposed \cite{cheng2023multimodal}, encompassing modality decoupling, unimodal knowledge retention, and multimodal knowledge retention. Its first module ensures that individual data modality in deleted pairs are treated as unrelated by the model, while the latter two focus on preserving essential knowledge to maintain the model performance.

\subsubsection{Federated learning.}
Federated learning enables collaborative model training without the need to share clients' local data, offering a more privacy-enhancing approach compared to centralized training methods.
The MMFed framework, proposed in~\cite{xiong2022unified}, addresses modality discrepancies in MMLS settings through a unified multimodal federated learning approach. It utilizes a cross-attention mechanism to identify key correlations across modalities on local clients, then uploads the parameters of both the classifier and attention module to a central server. 
CreamFL further advances multimodal federated learning by enabling larger server models and aggregation of clients with diverse architectures and data modalities, thereby aligning more closely with practical scenarios~\cite{yu2023multimodal}.

\subsection{Benchmark}
Current memorization measurements lack standardization and are primarily restricted to VLMs. Common metrics include \textit{Frechet Inception Distance score}~\cite{somepalli2023understanding} 
and \textit{Memorization score}~\cite{zhang2023forget}. 
Machine unlearning evaluation parallels that in unimodal settings, including attack-based methods like Membership Inference Robustness, and accuracy-based metrics like Mean Squared Error, alongside others like Activation Distance and Retraining Time~\cite{nguyen2022survey}.
Benchmarking multimodal federated learning is still in its early stages. 
FedMultimodal is the first systematic benchmark~\cite{feng2023fedmultimodal}, offering 10 representative datasets from 5 key MMLS applications
and providing a pipeline for effective multimodal federated learning assessment.

\subsection{Discussion}
Due to the diversity and richness of input data modalities, the potential risk of privacy leakage is concerning in MMLS.
This involves issues such as the recovery of removed sensitive information after multimodal fusions.
Besides, the increasing scale of MMLS further exacerbates the challenge of memorization.
Despite these concerns, privacy issues and their underlying causes are still not thoroughly explored, lacking formal conceptualizations and investigations. One potential direction is to extend findings in \cite{zhao2023unveiling} to multimodal data.
\section{Conclusion and Open Challenges}
This paper presents the first comprehensive survey on safe MMLS, introducing a systematic taxonomy centering on four critical dimensions: \textit{robustness, alignment, monitoring,} and \textit{controllability.}
Following this taxonomy, we review and summarize current research and progress. We hope that this survey offer insights for future research.
Below, we highlight several challenges in MMLS safety. 


\noindent\textbf{Memorization Problem in MMLS.}
Memorization in large-scale models is a significant contributor to privacy leakage. While this issue has been extensively explored in unimodal systems, research on memorization within MMLS remains limited. There is an urgent need for a formal definition of this problem and the establishment of unified metrics to measure it, especially given the recent rapid advancement of multimodal foundation models.

\noindent\textbf{Differentially Private Training.}
Differential privacy techniques demonstrate their effectiveness in safeguarding privacy in unimodal systems. Yet, adapting these techniques for MMLS faces challenges and is in its infancy. 
Some strategies to enhance differential privacy are to apply modality-specific noise addition and coordinate across modalities by considering the interdependence between modalities.


\noindent\textbf{Calibration and Uncertainty Quantification.}
Miscalibration in multimodal learning, a critical yet under-explored issue, poses challenges to MMLS safety. Future research should focus on understanding this phenomenon and developing advanced calibration methods. 
Besides, conformal prediction, known for its scalability and minimal assumptions, emerges as a promising solution for quantifying uncertainty.
Developing CP methods tailored explicitly for MMLS could significantly benefit practical applications.


\noindent\textbf{Advancing Datasets for MMLS Safety.}
Challenges with multimodal datasets include the scarcity of open-source large models capable of effectively handling multimodal inputs, limited interpretability of existing models that hinders understanding and trust, and the difficulty of quality control and data collection. In addition, research in MMLS is often restricted to specific use cases, and uneven dataset distribution hinders model generalization.
These issues require more efforts in data collection and refinement.

\noindent\textbf{Safety Challenges in Multimodal LLMs.}
Multimodal LLMs (e.g., GPT 4V~\cite{achiam2023gpt}) are designed to understand and generate content across different modalities. Safeguarding multimodal LLMs is very challenging due to various reasons including the additional safety rift induced by different data modalities, often black-box API access, unknown data sources, and emerging LLM safety issues such as hallucination, data leakage, and jailbreak. However, existing efforts to achieve safe multimodal LLMs have been scarce. 

\section*{Acknowledgements}
This material is based upon work supported by the National Science
Foundation (NSF) Grant \#2312862, \#2406648, \#2406647 and a Cisco gift grant.

\bibliographystyle{ACM-Reference-Format}
\bibliography{sample-base}
\end{document}